\documentclass[aps,twocolumn]{revtex4}
\usepackage{amsmath}
\usepackage{graphicx}
\usepackage{overpic}
\usepackage{xspace}

\newcommand{\gc}{g^{(2)}_{cc}(\tau)}
\newcommand{\gx}{g^{(2)}_{xx}(\tau)}

\begin{document}
\def\bra#1{\mathinner{\langle{#1}|}}
\def\ket#1{\mathinner{|{#1}\rangle}}
\def\braket#1{\mathinner{\langle{#1}\rangle}}
\def\Bra#1{\left<#1\right|}
\def\Ket#1{\left|#1\right>}

\title{Antibunching correlations in a strongly coupled exciton  - photonic crystal cavity system:
Role of off-resonant coupling
to multiple excitons}
\author{E. Illes}
\author{S. Hughes}
\affiliation{Department of Physics, Queen's University, Kingston, ON K7L 3N6 Canada}
\date{\today}

\begin{abstract}
We employ a master equation approach to study the
second-order quantum autocorrelation functions
 for up to two independent
 quantum dot excitons, coupled to an off-resonant cavity in a photonic crystal -
 single quantum dot system.
 For a single coupled off-resonant exciton, we observe novel oscillatory behavior in the early-time dynamics of the cavity  autocorrelation function,
  which leads to decreased antibunching  relative to the exciton mode.
 With  a second coupled exciton in the system,
 we find that the magnitude and the lifetime of these oscillations greatly increases,
  since the cavity is then able to exchange photons with multiple excitonic resonances.
   We unambiguously show that this spoils the antibunching characteristics of the cavity quasi-mode, while the autocorrelation of the first exciton is unaffected.
  We also examine the effects of detector time resolution and make a direct connection to
  a series of recent experiments.
\end{abstract}

\maketitle

{\em Introduction.--}
Two level behavior is ubiquitous in quantum physics,
and when a two-level system is coupled with a suitable optical cavity system,
the regime of cavity - quantum electrodynamics
(cavity-QED) can be realized.
Cavity-QED is a large and fascinating field, and several well known effects have been observed
and exploited for a number of years, e.g.,
the Purcell effect~\cite{purcell}  and
vacuum Rabi splitting~\cite{scully}.
Cavity-QED systems that 
emit {\em deterministic}  single photons
are also interesting 
from an applications viewpoint,
opening doors to practical ways of doing
quantum information processing~\cite{bennett, gisin2002}.
%
Two-level system behavior can occur in a number
of material structures, including
atoms~\cite{kuhn,kimble}, ions~\cite{keller},
molecules~\cite{brunel,lounis},  and color
centres~\cite{kurtsiefer}.
Semiconductor QDs~\cite{yuan,chang}
have also been proposed
as an ``artificial atom,''
with strong excitonic levels that mimic 
two level behavior and
are particularly attractive since they can function
as a scalable  and compact system for emitting
single photons.  Direct coupling between
QD excitons and semiconductor cavities has recently been
demonstrated in a variety of systems such as
micropillars~\cite{Moreau2001,Pelton2002,Reithmaier2004}, microdisks~\cite{Michler2000,Peter2005}, and  photonic crystal (PC) cavities~\cite{Yoshie2004,Hennessy:2007}.

An important difference between the modern QD semiconductor devices, and
earlier atomic systems, is that the confined photon cavity environment is considerably
more complicated than a couple of mirrors, frequently exploiting photonic bandgap physics and high-index-contrast
modes. In addition, QDs exhibit rich excitonic structure, hence it is not known to what extent
the two level approximation applies, though it is expected to be suitable for narrowband cavity
systems with a well defined cavity resonance that couples to the target exciton mode.
Thus, it can be anticipated that,
in addition to the familiar cavity-QED phenomena well known to
atoms, further complexity
unique to these semiconductor cavity - QD systems can been observed.
A recent example of  off-resonant coupling between a cavity and a single QD resulted in the observation of a
 pronounced
cavity mode emission~\cite{Hennessy:2007,Press:2007,Kaniber:2008}.
First considered a surprising result, it is now known that this is
likely due to the non-trivial coupling between the leaky cavity mode
and the exciton mode~\cite{Raymer2006,Auffeves2008,Yamaguchi2008,Naesby2008,Hughes:2009}, which is
exasperated in a planar PC system~\cite{Hughes:2009}.
Even more surprising has been the various reports of
poor antibunching - a measure of the quantum nature of the emitter -  even in the strong coupling regime for an exciton detuned from the cavity resonance~\cite{Hennessy:2007}; and  further experimental studies of the quantum correlation function in the off-resonant regime have provided additional perspectives on these results.
Specifically, it has been observed that the cavity autocorrelation function shows
 significantly worse antibunching than the exciton autocorrelation function~\cite{Press:2007},
 and to the best of our knowledge this perplexing behavior has not been yet
 explained. Yet an understanding of this mystery is a crucial step towards the
 understanding and design  of QD-based  single photon sources and
 would provide fresh modeling insight into the nanoscale  light-matter interactions
 at a very basic level.

In this Letter, we introduce a quantum optics formalism that allows one to calculate both the
  exciton and the cavity mode quantum autocorrelation functions for a single QD - PC cavity system.
  We provide a framework in which to understand the above experimental results,
  by emphasizing the consequences of the quasi-mode nature of the cavity, and the significant effect of additional excitons on the quantum autocorrelation functions of the system.
  We show the general need to account for coupling to additional excitons in the system
  that play a qualitatively important role, even if they are far off resonance. We investigate the nature of
  off-resonant coupling and pure dephasing in detail, and make a direct and successful connection
  to recent experimental results. 

\begin{figure}[b!]
\centering
 \includegraphics[width=7.5cm]{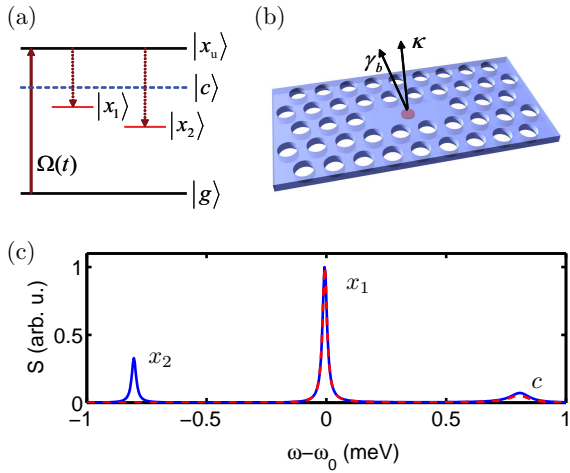}
\caption{(Color online) System under investigation. (a) Energy level diagram for theoretical model.  (b) Photonic crystal slab with embedded quantum dot, showing a background radiative
 decay
 process and a leaky cavity decay process.
 (c) Cavity mode emission spectra for one (dashed red) and two (solid blue) excitons.}
\label{fig:system}
\end{figure}

{\em Theory and Model.--}
Our model system  consists of a
single QD embedded in a planar PC cavity, but is also applicable to other
 semiconductor cavity systems where the leaky cavity mode dominates the emission, e.g.,
the micropillar cavities. Due to the leaky nature of the cavity mode, both the exciton and the cavity are able to emit photons out of the plane (vertically).  In Fig.~\ref{fig:system}(a), we show the relevant energy level diagram, that consists of up to two excitons (both originating from a single QD and assumed to be independent -- though they
are indirectly coupled through the cavity mode) and the resonant mode of the PC cavity.  The uppermost exciton energy level is utilized for exciting the system
with a far off-resonant pump pulse.
We adopt a master equation approach~\cite{Carmichael:1989} to calculate the second-order quantum correlation functions.
The master equation for the leaky cavity and the radiative decay modes, in the interaction picture, is 
\begin{align}
  \dot{\hat \rho} =&\frac{1}{i\hbar}[\hat H_I,\hat \rho]
  -\kappa(\hat a^{\dagger}_c\hat a_c\hat\rho+\hat\rho \hat a_c^{\dagger}\hat a_c-2\hat a_c\hat\rho \hat a^{\dagger}_c)
  \nonumber \\
&  \!\!\!\!\!\!\!\!\!\! -\!\sum_{i=1,2} \! \left [ \gamma_i(\hat \sigma_{ii} \hat \rho
+\hat \rho \hat \sigma_{ii}  - 2\hat \sigma^- \hat \rho \hat \sigma^+) \!-\!\gamma^\prime_i (\hat \sigma_{ii} \hat \rho \hat \sigma_{00} + \hat\sigma_{00}\hat\rho \hat\sigma_{ii}) \right ]\nonumber \\
& \!\!\!\!\!\!\!\!\!\!  - \sum_{i=1,2} \! \gamma_{iu}(\hat \sigma_{uu}\hat \rho+\hat \rho\hat \sigma_{uu}-
2\hat \sigma_{iu}\hat \rho\hat\sigma_{ui}),
\end{align}
with the interaction Hamiltonian
\begin{align}
\!\!\!\!\! \hat H_I=i\hbar \!\sum_{i=1,2}g_i(\hat\sigma^{-}_i\hat a^{\dagger}_c-\hat \sigma^{+}_i\hat a_c)
 +\frac{i\hbar\Omega(t)}{2} \left (\sigma^+_u -\sigma^-_u\right ),
\end{align}
where $\Omega(t)$ is a classical excitation pulse that excites
the upper lying exciton level, $\ket{x_u}$,
$2\gamma_{iu}$ is the fast non-radiative decay rate,
$\hat \sigma_{ii}=\hat\sigma^+_i\hat\sigma^-_i$, and
$\hat \rho$ is the density matrix operator.  The boson operators $\hat a_c$ and $\hat a^{\dagger}_c$ are the photon creation and annihilation operators for the cavity mode, and $\hat \sigma^{+}_i$ and $\hat \sigma^{-}_i$ are the Pauli raising and lowering operators, respectively.
 The  parameters of the system include $g$ - the coupling strength between the cavity and exciton; and $2\gamma_i$, $2\kappa$ and $2\gamma\prime_i$ - the radiative, cavity and pure dephasing decay rates, respectively.

Importantly, both 
the one and two time equations of motion can be calculated from the master equation~\cite{Carmichael:1989}, and utilizing the
 quantum regression theorem~\cite{scully}, we
 obtain the second-order quantum correlation functions
\begin{equation}
\begin{split}
G^{(2)}_{x_i,x_i}(t,t+\tau)\propto
\braket{\hat \sigma^+_i(t)\hat \sigma^+_i(t+\tau)\hat \sigma^-_i(t+\tau)\hat \sigma^-_i(t)}\\
G^{(2)}_{c,c}(t,t+\tau)\propto
\braket{\hat a^{\dagger}_c(t)\hat a^{\dagger}_c(t+\tau)\hat a_c(t+\tau)\hat a_c(t)},
\end{split}
\end{equation}
for the exciton and the cavity mode, respectively.
The time integrated $g^{(2)}(\tau)$ is then obtained
by integrating over time $t$ for the center pulse, and normalizing
to the  peak of one of the sideband pulses~\cite{Kiraz:2004}.

\begin{figure}[b!]
\centering
\includegraphics[width=\columnwidth]{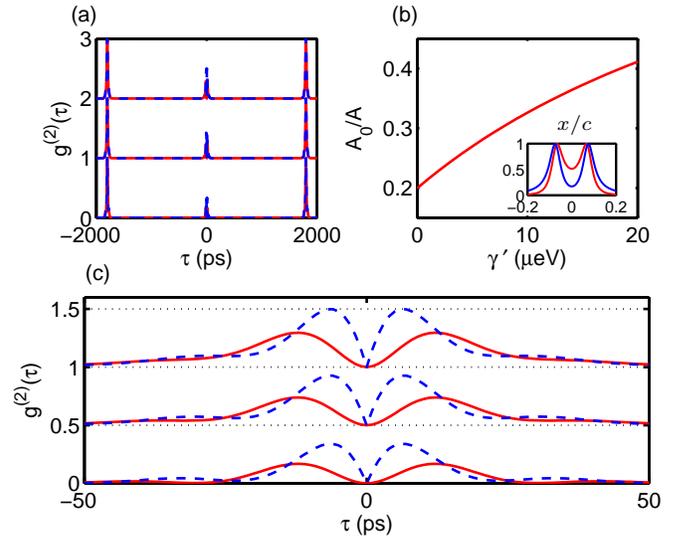}
\caption{(Color online) (a) Autocorrelation function for the cavity (solid red) and exciton (dashed blue) for pure dephasing rates of $\gamma\prime= 0,\,10$ and $20\,\mu$eV from bottom to top;
  $\gamma=\,1\mu$eV in all cases. (c) Zoom about $\tau=0$ region of (a). (b) Degree of antibunching - the ratio of the area under the $\tau=0$ peak compared to peaks at other times. Inset: Associated spectrum - for an exciton on resonance with the cavity in the strong coupling regime.}
\label{fig:choose_gamma_prime}
\end{figure}

{\em Calculated Spectra and Antibunching Correlations.--}
 We optically excite the system up to a higher-lying level ($x_u$) from the ground state via Gaussian excitation pulses, $\Omega(t)$, with full width at half-max of 10\,ps spaced 1800\,ps apart;   the pulse amplitudes correspond closely to $\pi$-pulses to optimally feed the high-lying level.  The system is then allowed to quickly relax from $\ket{x_u}$ to 
 $\ket{x_i}$ via a fast  non-radiative decay rate of $2\gamma_{iu}=0.4\,$meV.  The cavity emission spectra calculated in the presence of one and two excitons can be found in the lower panel of Fig.~\ref{fig:system}.  In the red (solid line) spectrum, a single exciton couples to the leaky cavity mode and the cavity emission spectrum consists of both the bare exciton resonance as well as the bare cavity resonance. These one exciton results are
  consistent with the results of others~\cite{Hughes:2009,Auffeves2008,Yamaguchi2008,Naesby2008}.
Also,  
  since the cavity mode
  emission characteristics entirely dominate the overall collected spectrum~\cite{Hughes:2009},
  we only show the cavity mode emission in the spectrum.
  As recognized, introducing a second exciton has  a small but non-negligible effect on the spectrum.
   Analogously to the first exciton, the second one also ``feeds'' the cavity mode, and therefore has the effect of increasing the weight under the peak at the bare cavity resonance as well as introducing a peak at its own bare resonance.  Again, we emphasize that the cavity emission spectrum will dominate the emission characteristics, and contains resonances at both the bare cavity resonance as well as the resonances of the excitonic modes that {\em feed} it.  

\begin{figure}[b!]
\centering
\includegraphics[width=8cm]{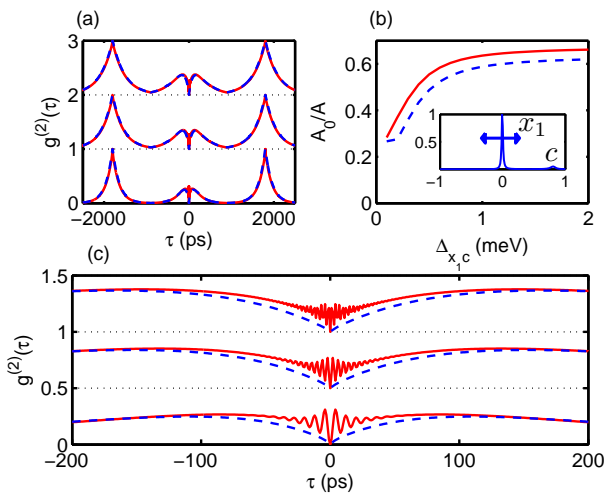}

\caption{(Color online) (a),(c) Cavity (solid red curve) and exciton (dashed blue curve) autocorrelation functions with a zoom in on the region near $\tau=0$ in the lower panel and the long time behavior in the upper panel.  The detuning from bottom to top are $\Delta_{xc}= 0.4, 0.8$ and $1.2$ meV.  (b) Antibunching dip for the cavity (solid red) and the exciton (dashed blue). Inset: cavity emission spectrum schematic. }
\label{fig:move_dxc1}
\end{figure}

In choosing model parameters, 
we aim to mirror those determined from
related experiments~\cite{Hennessy:2007,Press:2007,Kaniber:2008}, and choose a cavity-exciton coupling of $g=0.075$ meV, and half rates of radiative and cavity decay of $\gamma=1\,\mu$eV and $\kappa=0.05$ meV, respectively.  To determine an appropriate pure dephasing rate, we calculate the rate from a representative amount of antibunching for an on-resonant exciton and varying $\gamma\prime$.  We then compare our results to those typically seen in experiment~\cite{Press:2007,Hennessy:2007}.  The second order correlation functions are shown in Figs.~\ref{fig:choose_gamma_prime}(a) and (c).  The theory utilized is amenable to projecting onto either the cavity quasi-mode or the excitonic mode separately - even though the system is on resonance - which is something that is not experimentally accessible.  However, since the cavity mode dominates the emission,
the on-resonance condition is
heavily dominated by the cavity mode contribution.
With the chosen parameters, we
extract a representative
pure dephasing rate of $\gamma\prime= 2\,\mu$eV - which corresponds to an antibunching dip of 23\% on resonance (see Fig.~\ref{fig:choose_gamma_prime} (b)).

We now introduce a detuning between the exciton  and the cavity resonance $\Delta_{xc}=\omega_c-\omega_x$, where $\omega_c$ and $\omega_x$ are the bare cavity and exciton resonances, respectively.  We examine a system with a singly coupled, off-resonance exciton
$x=x_1$, as well as an upper loading exciton, $x_u$.  The calculated
autocorrelation functions are shown in Fig.~\ref{fig:move_dxc1} for three different detunings.  In general, we observe an overall broadening of all peaks (as compared to the resonant case) and a reduction in antibunching.  This reduction is magnified with increased detuning.  However, for all values of detuning seen in Fig. \ref{fig:choose_gamma_prime} (b), we note that $\gc$ shows decreased antibunching relative to $\gx$, which agrees with the experimental observations reported by Press {\em et al.}~\cite{Press:2007}.  For larger detuning, the size of the antibunching dip levels off to about 62\% and 67\% for the exciton and the cavity, respectively.  We note that the exciton is expected to experience some antibunching even in the absence of the cavity~\cite{Kiraz:2004},
subject to the amount of pure dephasing, which would explain the limiting case seen for large detuning.
A closer examination of the early time dynamics of the autocorrelation functions for a single exciton (see Fig. \ref{fig:move_dxc1} (c)) reveals new oscillations in the $\gc$ that are not seen for $\gx$.  These oscillations are caused by the off-resonant coupling of the cavity to the exciton, and are manifestations of the quasi-mode nature of the cavity mode.  The period of the oscillations corresponds to the detuning frequency of the exciton $T_{\rm osc}\propto1/f_{\rm detune}$, and therefore the frequency of the oscillations increases with increased detuning.

%
%

\begin{figure}[t!]
\centering
\includegraphics[width=8.5cm]{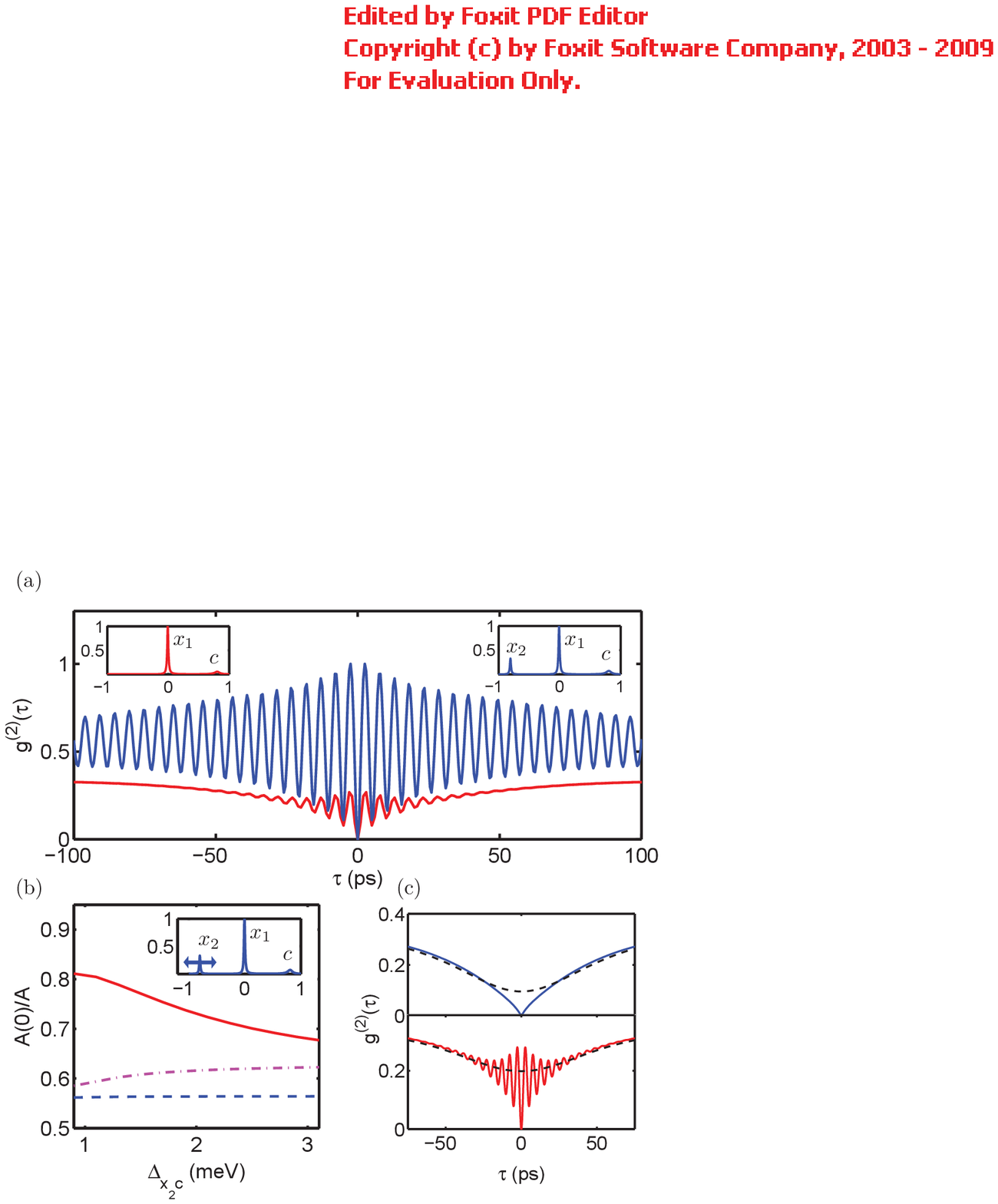}



\caption{(Color online) (a) Cavity autocorrelation functions for one (smaller oscillations) and two (larger oscillations) excitons.  The detunings for the two excitons are $\Delta_{xc}=0.8\,$meV and $\Delta_{xc}=1.6\,$meV.  Inset: One (red) and two (blue) exciton spectra showing cavity mode emissions.  (b) Antibunching for the cavity mode (solid red), first excitonic mode (dashed blue), and second excitonic mode (dot dashed magenta).  The detuning of the first exciton is constant at $\Delta_{x_1c}=0.8$\,meV, while $\Delta_{x_2c}$ varies.  Inset: Cavity emission spectrum schematic.  (c) The effect of finite detector time resolution on $\gx$ (top) and $\gx$ (bottom) shown by a dashed black line.  These were calculated for a single exciton that is detuned by $\Delta_{x_1c}=0.8$\,meV from the cavity.}
\label{fig:one_to_two}
\end{figure}

We next turn to describing the effects of other nearby excitons on the ensuing quantum statistics.  We include a second exciton that is essentially identical to the initial one (same $\gamma,\gamma\prime,g$) with the exception of its detuning with respect to the cavity mode, $\Delta_{x_2c}$.  We show a representative plot of $\gc$ in the presence of one versus two excitons coupled to the cavity in
Fig.~\ref{fig:one_to_two}~(a).  For the case shown, the second exciton is doubly detuned from the cavity relative to the first exciton.
In fact, in experiments, the coupling strength of the
second exciton may be significantly larger than the
first one (eg. Ref.~\cite{Hennessy:2007}) which will only exasperate the results that follow.
As can be seen in Fig.~\ref{fig:one_to_two}~(a), the presence of the additional exciton has
profound
influence on $\gc$.  These extraneous effects consist of magnifying all the features that were previously unique to $\gc$.  In particular, we find that both the magnitude, as well as the lifetime of oscillations in the early time dynamics are significantly increased - thereby magnifying the differences in the antibunching of the cavity versus the exciton mode.

In Fig.~\ref{fig:one_to_two}~(b), we examine the effects of varying the detuning of the second exciton and summarize the results on antibunching.  In this figure, we maintain an equal coupling strength of $g=0.075$ meV for both excitons, and note that the oscillator strength of the second exciton line in the cavity emission spectrum decreases with increased detuning.  We find that the antibunching of the stationary exciton ($x_1$) is unaffected by the detuning of the other exciton - which is expected since we assumed no direct coupling between the excitons.  For the second exciton ($x_2$), we find that increasing the detuning from the cavity resonance spoils the antibunching of the exciton, though it still retains a significantly stronger antibunching dip than the cavity mode.  For the cavity mode, we find that in the presence of nearby additional excitons, the antibunching is largely destroyed, particularly relative to that of the excitonic modes.  However, as the second exciton is moved further off resonance the antibunching of the cavity mode improves, and for large $\Delta_{x_2c}$ we begin to recover the single exciton case.  In cases where the second excitonic line is more intense than the initial one
or with more excitons excited (e.g.,~\cite{Hennessy:2007}), we find that the substantial
 difference between the antibunching of the excitons and the cavity mode increases even further,
 again consistent with experimental observations.

Finally, in order to connect more closely with experimental detectors, we include the effects of finite detector time resolution, by modeling the detector time response as a Gaussian function with a full-width at half-max of 50 ps and an area of one.  We apply this averaging to the output of the $\gc$ and $\gx$ calculations and obtain the results seen in Fig.~4(c).  The primary effect of finite detector response is an averaging out of any behavior that occurs at time scales shorter than the 50\,ps response time, such as the quick drop to zero seen on both the cavity and exciton autocorrelation for $\tau\rightarrow 0$, and the oscillatory behavior observed in $\gc$.  However, this averaging preserves the relative areas under the peaks in the autocorrelations, and therefore has no effect on the size of the antibunching dips.

{\em Conclusion.--}
By utilizing a master equation model to describe the coupling between a leaky
cavity mode and one or more QD excitons,
we find significant differences between the second-order quantum
autocorrelation functions of the
exciton modes and cavity modes.
These differences are manifestations of the quasi-mode nature of the cavity and include oscillations in the early time dynamics of $\gc$, as well as the reduced antibunching of the cavity mode relative to the exciton.  Furthermore, we find that the presence of additional excitons magnifies this difference
and largely destroys the antibunching of the cavity mode, which qualitatively explains recent, and hitherto
unexplained, experimental observations~\cite{Hennessy:2007,Press:2007}.


{\em Acknowledgments.--}
This work was supported by the National Sciences and Engineering Research Council of Canada, the Canadian Foundation for Innovation, and the Walter C. Sumner Memorial Foundation.




\end{document}